\documentclass[twocolumn,apj]{emulateapj}

\begin{document}

\title{Unusual filaments inside the umbra}


\author{L. Kleint$^1$ and A. Sainz Dalda$^2$}
\affil{$^1$High Altitude Observatory / NCAR, P. O. Box 3000, Boulder, CO 80307, USA; kleintl@ucar.edu\\
$^2$Stanford-Lockheed Institute for Space Research, Stanford University, 
HEPL, 466 Via Ortega, Stanford, CA 94305, USA}




\begin{abstract}
We analyze several unusual filamentary structures, which appeared in the umbra of one of the sunspots in AR 11302. They do not resemble typical light bridges, neither in morphology, nor in evolution. We analyze data from SDO/HMI to investigate their temporal evolution, Hinode/SP for photospheric inversions, IBIS for chromospheric imaging and SDO/AIA for the overlying corona.
Photospheric inversions reveal a horizontal, inverse Evershed flow along these structures, which we call umbral filaments. Chromospheric images show brightenings and energy dissipation, while coronal images indicate that bright coronal loops seem to end in these umbral filaments. These rapidly evolving features do not seem to be common, and are possibly related to the high flare-productivity of the active region. Their analysis could help to understand the complex evolution of active regions.

 \end{abstract}


\section{Introduction}

Bright structures can often be observed inside the umbra, either as umbral dots or light bridges (LBs). While the term  ``light bridge" is often used to denote a bright lane separating a sunspot's umbra into two parts, several classifications were proposed, which  also included ``penumbral'' or ``faint" LB that are only intrusions into the umbra \citep{vazquez1973,sobotka1997}. Most LBs show a sequence of granules, similar to photospheric granulation, or aligned bright grains. Very few seem to have a filamentary structure, looking as if penumbral filaments extend into the umbra. Indeed, before sub-arcsecond resolution images were available it was thought that ``a granular structure is common in all light-bridges'' \citep{vazquez1973}. To our knowledge, the first picture of a filamentary structure in the umbra was shown by \citet{muller1979}, where it seems that two sets of several penumbral filaments protrude into the umbra and later connect. \citet{livingston1991} reported an observation where a filament seemed to continue into the umbra from the penumbra and he speculated that the field may be more inclined there. But the resolution of the photograph 
is too limited to resolve structures inside the LB. None of these filaments however are reported to appear suddenly or to change their propagation direction, both of which we will present in this paper.

The origin and evolution of LBs is still unclear. Their magnetic field strength is lower than the surrounding umbral field and their inclination is more horizontal \citep{beckersschroeter1969, litesetal1991, ruedietal1995, leka1997}.  Chromospheric activity was reported above LB in the form of plasma ejections and localized brightenings \citep{roy1973,asaietal2001,louisetal2008, shimizu2011, liu2012}. \citet{louisetal2009} found strong photospheric downflows of up to 10 km/s in LB, some of which are co-spatial with simultaneously occurring chromospheric brightenings. Because of the different inclination and field strength of the LB compared to the umbral field, currents or stress, possibly leading to reconnection, may be expected.

In this paper, we will present observations of some unusual LB, or rather filaments protruding into the umbra, which occurred in one spot of the very flare-productive AR 11302. They do not resemble the typical granular LB, they do not cross the umbra, nor do they lead to the splitting and decay of this sunspot. Instead, they appear within hours and look as if a penumbral filament of more than 10\arcsec\ length was launched into the umbra. The only similar observation we found in the literature can be seen in Fig.~1 of  \citet{ruedietal1995} showing LB in the shapes of hooks, which are no longer visible on the next day's image, suggesting that they are an equally dynamic phenomenon as our observations. Our chromospheric images show bright plasma above these filaments, which appear to be connected in the chromosphere and seem to correspond to footpoints of coronal loops.


\section{Observations}
\label{obs}

AR 11302 was visible to Earth-bound observations from September 22, 2011 -- October 4, 2011. For our analysis, we combine data taken by Hinode (SOT/SP), SDO (AIA and HMI) and IBIS.

\subsection{Hinode SOT/SP}
SOT/SP aboard the Japanese \textit{Hinode} satellite \citep{kosugietal2007, tsuneta2008sot, suematsuetal2008, ichimotoetal2008} is a slit-scanning spectropolarimeter with a spatial resolution of 0.32\arcsec, a spectral resolution of 21.5 m\AA\ and a polarimetric sensitivity of 10$^{-3}$ I$_c$. We analyzed 10 maps, taken between 23 September 2011, 23:21 UT and 26 September 2011, 11:02 UT. The data reduction was performed with the standard \textit{SolarSoft} routines in IDL, mainly developed by B. Lites.

The strength of SOT/SP is a high-resolution wavelength coverage for both the 6301.5~\AA\ and the 6302.5~\AA\ lines. These data are ideally suited for inversions to derive photospheric atmospheric parameters. The disadvantage of SP is the required duration to scan across the solar surface to obtain one map (about 40 minutes for the analyzed observations). Therefore, other instruments were used to analyze dynamic phenomena.

\begin{table*}
\caption{Properties of the unusual filaments.}\label{tab1}
\begin{tabular}{l | l l c c c c}
\hline
\centering
                   & life time [date] & [h] & length [\arcsec] & width at umbra [\arcsec] & width at ``middle" [\arcsec]& width at end [\arcsec]\\
\hline\hline
UF~1 & Sep 23, 20:00 -- Sep 26, 15:00 & 67 & 12.2 & 1.1 & 0.90 & 0.74 \\
UF~2 & Sep 24, 14:45 -- Sep 24, 23:15 & 8.5 & 10.5 & 1.1 & 0.85 & 0.59 \\
UF~3 & Sep 24, 1:15 -- Sep 25, 4:00   & 26.75 & 13.0 & 1.4 & 1.25 & 0.84 \\
\end{tabular}
\end{table*}

\subsection{SDO}
The Solar Dynamics Observatory \citep[SDO,][]{pesnelletal2012} is a satellite with two different imaging instruments: AIA \citep{lemenetal2012} and HMI \citep{scherreretal2012,schouetal2012}.

We analyzed AIA images in the wavelengths of 4500~\AA, 1600~\AA, 304~\AA\ and 171~\AA. HMI was used to obtain more frequent continuum images, as AIA only takes one frame per hour in 4500~\AA. AIA and HMI data were reduced with \textit{SolarSoft}, which despikes, flat fields, removes bad pixels, rotates to a common coordinate system and interpolates all data onto a 0.6\arcsec/px plate scale.

AIA's strength are continuous images in many wavelength regions, apart from short blackouts during eclipses that lasted about 1 hr on our observing dates. A disadvantage is the lower spatial (AIA and HMI) and spectral (HMI) resolution, which for example did not show the complex Stokes profiles in the observed filaments.

\subsection{IBIS}
IBIS is a dual Fabry-Perot imaging spectropolarimeter \citep{cavallini2006, reardoncavallini2008}. Several spectral lines with user-defined spectral steps can be scanned in sequence. IBIS has an approximate field of view (FOV) of $40\arcsec\ \times 80$\arcsec\ in polarimetric mode, a spatial resolution of 0.2\arcsec, and a polarimetric sensitivity of 10$^{-2}$.

We recorded 20 scan sequences of the western spot of AR 11302 on Sep 24, 2011, from 17.51 UT to 18.22 UT. One scan of the \ion{Ca}{2} 8542 \AA\ line contained 23 wavelength points with a spectral sampling of 70 m\AA\ around the line core and lasted 25 s. Three other spectral lines (\ion{Fe}{1} 6302 \AA, H $\alpha$ 6563 \AA\ and \ion{Na}{1} 5896 \AA) were also scanned, but are not presented in this paper. The seeing was very variable, making it necessary to exclude sequences where the seeing was not constant and good during the 25 s scan time. IBIS records six modulation states per wavelength (called $I+Q$, $I-Q$, $I+U$, $I-U$, $I+V$, $I-V$, although each state is a linear combination of all Stokes parameters), which after demodulation result in one image of the Stokes vector ($I$, $Q$, $U$, $V$).

The data reduction for IBIS includes a dark correction, flatfielding and alignment of the broadband and the narrowband channels, the first of which is speckle-reconstructed with KISIP \citep{woegervdl2008} and used to destretch the images of other channel. A disadvantage of the collimated mount of the Fabry-Perots is the wavelength-variation across the FOV. The final images are interpolated to a common wavelength scale to obtain truly monochromatic images. A polarization calibration taking into account the telescope and the instrument properties is also applied. All these steps can be done nearly automatically with a GUI that we developed. Seeing variations during the scan will lead to a bad wavelength interpolation, and seeing variations during the six modulation state images will lead to crosstalk. Therefore, we completely omitted scans with variable seeing from the analysis.

The main advantage of IBIS is the capability for imaging spectropolarimetry in the chromosphere. Being the only ground-base instrument used for this study, its main disadvantage is the variable seeing and the limitation of observing time.

\section{Unusual filaments}

  \begin{figure} 
   \centering
    \includegraphics[width=0.5\textwidth]{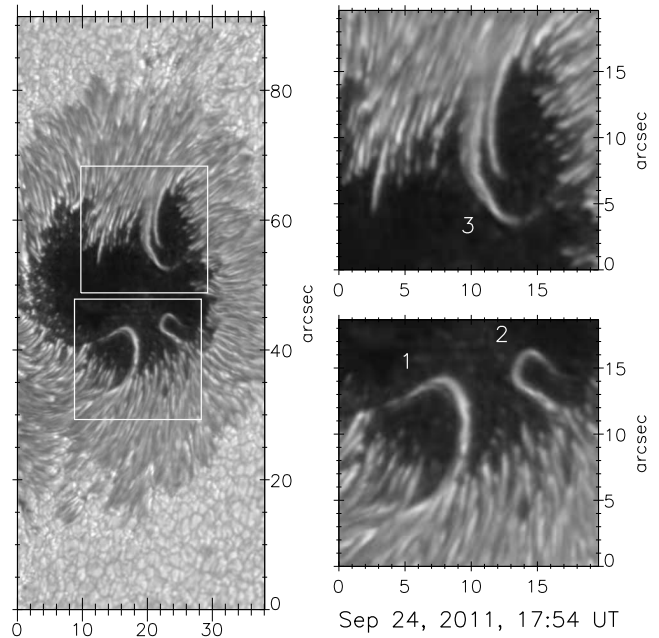}
   \caption{Speckle reconstructed IBIS whitelight image showing the three unusual filaments. The numbers correspond to the numbering in Table~\ref{tab1}.
                 }
         \label{filam}
   \end{figure}

During the evolution of AR 11302 at least three unusual filaments could be observed and are shown in Figure \ref{filam}. We will use the word ``filament'' in the photospheric sense, as in ``penumbral filament'', but these observations are not related to the chromospheric, large-scale filaments seen in H $\alpha$. All of our observed filaments were wider than regular penumbral filaments and moved inside the umbra. Table~\ref{tab1} summarized their properties, such as life times and sizes. The life times are approximate because the spatial resolution of SDO does not allow to determine the exact appearance and disappearance. Filaments, which occurred too close to the solar limb were omitted from the analysis. The width of the filaments was measured in three locations in Fig.~\ref{filam}, once at the edge of the umbra, once at their approximate middle and once where the filament seems to end, which is the least exact value because faint traces may be seen further out or the filament head splits.


\begin{figure*} 
   \centering
    \includegraphics[width=.97\textwidth]{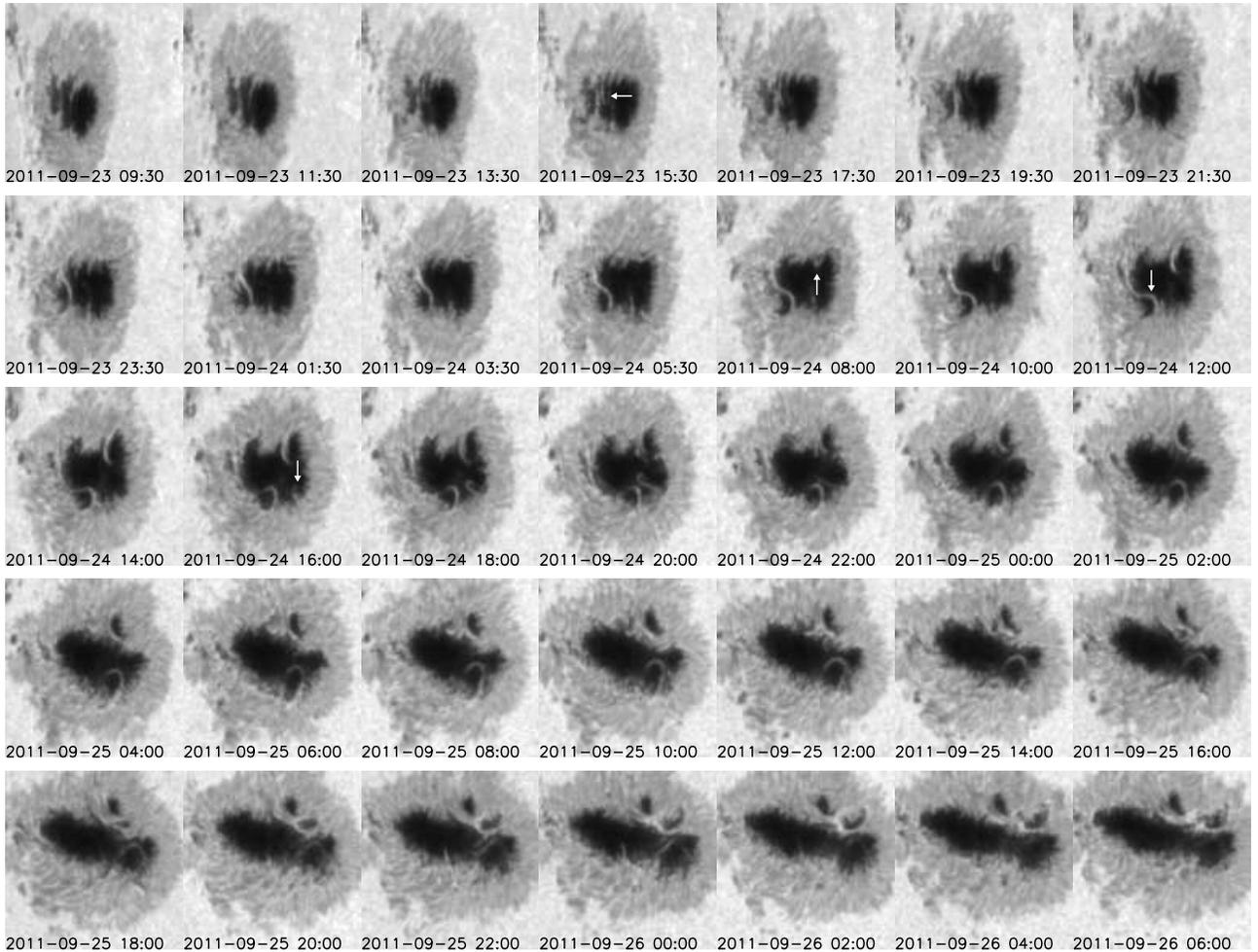}
   \caption{Temporal evolution of the umbral filaments (denoted by arrows) in 2 hr steps during about 3 days. These continuum images were taken by SDO/HMI, which was eclipsed by the Earth around 7 am each day, which is why some frames are missing.
                 }
         \label{temp_filam}
   \end{figure*}

Bright penumbral filaments have a typical width of 0.3--0.6\arcsec\ and a life time of few minutes to 4 hr \citep[][and references therein]{solanki2003}. Newer measurements \citep{rouppeetal2004} do not find any preferred width of penumbral filaments, but instead conclude that unresolved filaments with a width below 0.1\arcsec\ are present. Our observed filaments have about double or triple width of typical bright penumbral filaments and their life times are much longer. Their average intensities are equal to those of penumbral filaments. Another difference seems to be that while penumbral filaments often have a very bright (brighter than continuum intensity) head, our filaments become fainter towards their heads.

Because these filaments do not have the same properties as typical penumbral filaments, could they be more similar to LB? Our observed filaments do not seem to have a granular structure, ruling out a similarity to most LB. Also, they do not separate umbrae and all of them are hook-shaped, which is not typical for LB. Because there are so many definitions for LB, spanning different sizes (from below arcsec to several arcsec) and morphologies (granular, filamentary) one might argue that they are a rare, special type of LB. But because they do not resemble most LB and in fact, do not ``bridge'' across the umbra, we will call them \textit{umbral filaments} (UF).

\subsection{Temporal Evolution} 
\begin{figure*}
   \centering
    \includegraphics[width=.6\textwidth]{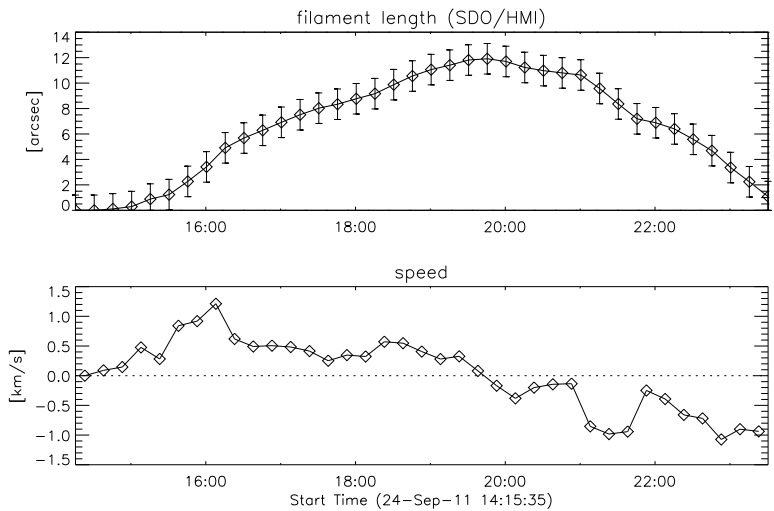}
   \caption{Upper panel: length of UF 2, measured from SDO/HMI continuum images. The error bars denote an uncertainty of $\pm$ 2 px for the manual length determination.
      Lower panel: apparent change of the filament's length per 15 minutes, smoothed over 45 minutes to minimize errors.
                 }
         \label{dynfilam}
   \end{figure*}

\begin{figure*}
   \centering
    \includegraphics[width=\textwidth]{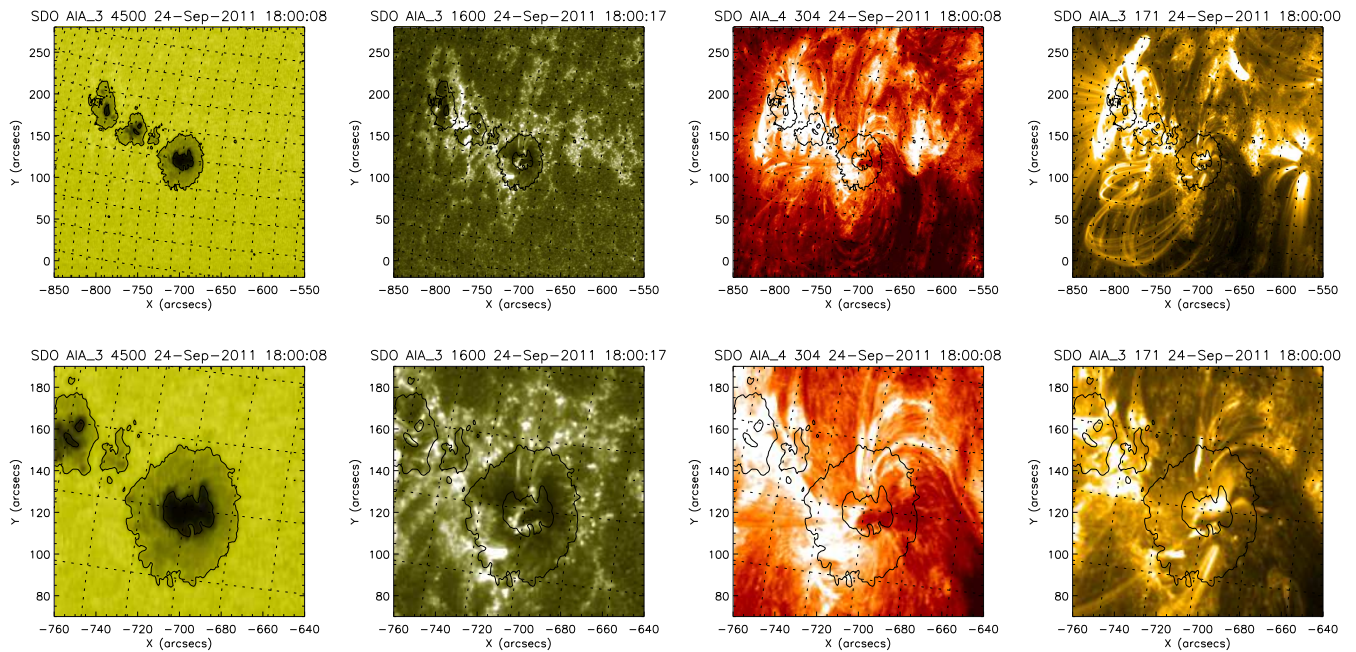} 
   \caption{SDO/AIA images of AR 11302 in the wavelengths 4500 \AA, 1600 \AA, 304 \AA\ and 171 \AA\ (left to right), going upwards in the solar atmosphere. All images were taken within seconds of 18:00 UT on Sep 24, 2011, close to the time of Fig.~\ref{filam}. The upper panel shows a FOV of 5\arcmin $\times$ 5\arcmin, and the lower panel 2\arcmin $\times$ 2\arcmin. Some coronal loops seem to end near or inside the UFs.
                 }
         \label{aia}
   \end{figure*}

Figure~\ref{temp_filam} shows the temporal evolution of the western spot of AR 11302 during about 3 days. The white arrows point out UFs. The NE filament (arrow at 2011-09-23, 15:30) is questionable because of the sunspot's foreshortening due to the proximity of the limb. UF 1 (2011-09-24, 12:00) seems to evolve from a LB,  whose north-eastern half merges with the penumbra, leaving only the hooked filament visible. The UF seems to be formed from the filamentary looking end of the LB, while the rest of the LB showed a granular structure (see Fig.~\ref{invoverview} below for high-resolution intensity image). It vanishes around 15:00 on Sep 26 by losing contrast compared to the umbra. The sunspot, or at least the filament, seems to rotate during its lifespan.

UF 2 (2011-09-24, 16:00) is the most dynamic filament with a life time of 8.5 hr. Its length and change of length can be seen in Fig.~\ref{dynfilam}. The length was measured from SDO/HMI images in intervals of 15 minutes from the edge of the penumbra, along the filament, to the last visible pixel of the filament head. Because of the slightly lower spatial resolution of SDO, the filament length is underestimated by about 1.5\arcsec\ compared to IBIS in Fig.~\ref{filam}, where its faint head can be seen. We estimate an additional error of $\pm$ 2 pixels arising from the uncertainty where the filament ends in the SDO/HMI images and the error bars only reflect this error. The measured values were smoothed in an interval of 45 minutes to minimize that measurement inaccuracies influence the velocity calculation. The apparent change rate lies between 1.2 and --1.1 km/s, which is the same order of magnitude of velocities of umbral dots, penumbral grains or the average Evershed flow \citep[][and references therein]{delT00,stix2002,solanki2003}.

UF~3 (2011-09-24, 8:00) appeared as a projected loop, whose lower half vanished after a few hours. Its projection angle is different than that of UF~1, which looks more horizontal.
Over the next 24 hr, UF~3 curled westwards, creating a temporary LB and a channel of opposite polarity (shown in Section~\ref{inversions}). The small spot created by this LB vanished later on Sep 28 and the light bridge became indistinguishable from the penumbra.

\section{Corona above umbral filaments}
  \begin{figure*} 
   \centering
    \includegraphics[width=0.6\textwidth]{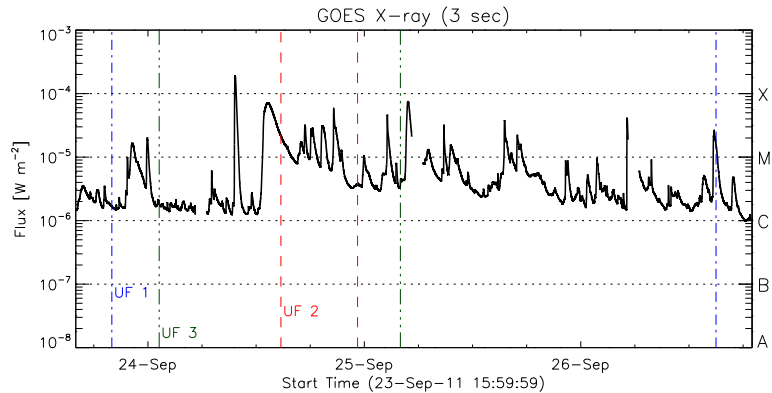}
   \caption{GOES X-ray flux during part of the evolution of AR11302 with the times when the UFs appeared and disappeared indicated by vertical dashed lines.
                 }
         \label{goes}
   \end{figure*}
   
  \begin{figure*} 
   \centering
    \includegraphics[width=\textwidth]{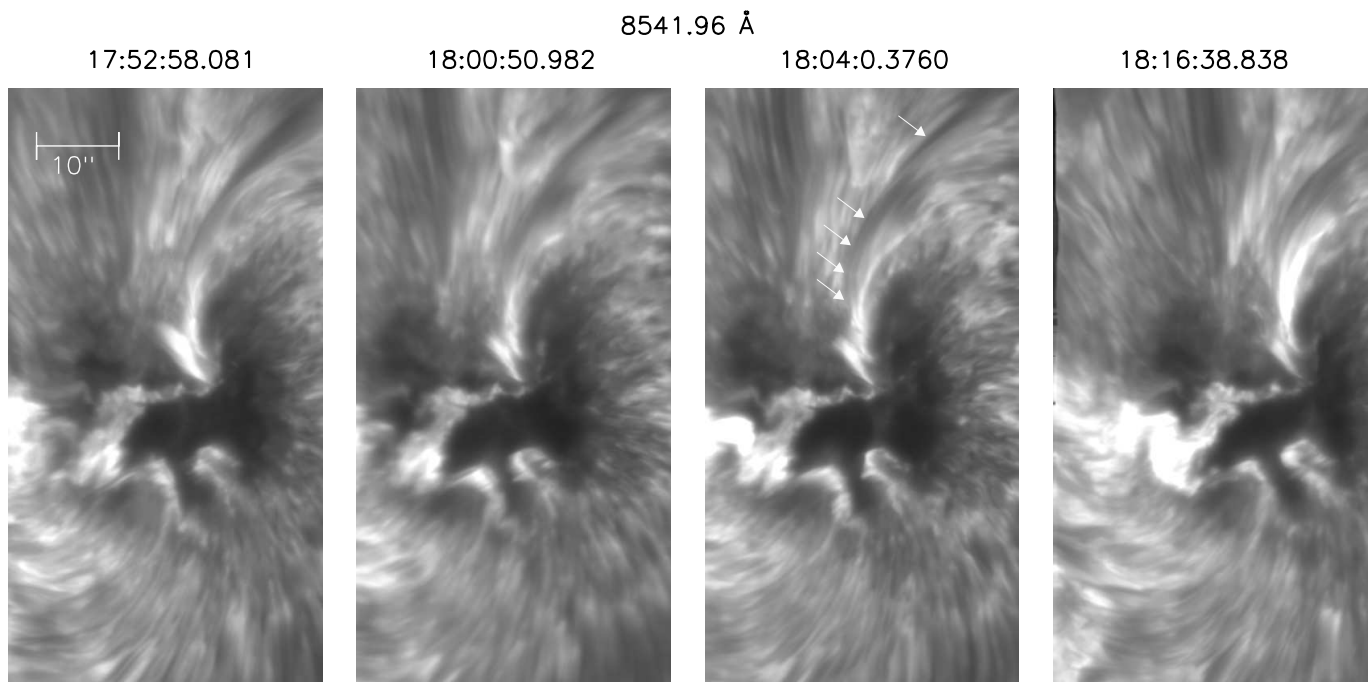}
   \caption{IBIS intensity images close to the line core position of the 8542 \AA\ line, taken on Sep 24, 2011. UF~1 and 3 are connected in the chromosphere and brighten after a flare occurs. The FOV of the last image is slightly shifted and shows a larger part of one of the flare footpoints. 
                 }
         \label{ibis8542}
   \end{figure*}

SDO/AIA images allow us to study the overlying corona. Figure~\ref{aia} shows emission from four different height ranges in the solar atmosphere, from the photosphere to the corona (left to right). Contours of the active region are drawn on each image, representing 0.9 and 0.3 I$_c$. The lower panels are a magnification of the upper panels, centered on the spot with the UFs. The 171~\AA\ panels show one loop system south of the active region and its western footpoints seem to end in the two southern UFs and their extension in the penumbra. 
Enhanced emission can be seen in 1600~\AA\ at these places. The northern filament seems to be the footpoint of another loop system, which continues towards the west and is best visible in 304~\AA.

The visibility of these loop systems was very variable and they changed rapidly as many strong flares occurred on Sep 24, 2011. Most of the strong (M and larger) flares took place in AR 11302. The X-ray flux while the UFs were present is shown in Fig.~\ref{goes}. The color-coded vertical lines denote the appearance and disappearance of the three observed filaments. Because these filaments seem to be related to coronal loops and loops change significantly during flares, there probably is some correlation between flares and UFs. It is however unclear if for example shearing motions in the photosphere (filaments moving) cause the coronal loops to rearrange, reconnect and flare. However, it is interesting that the most dynamic UF 2 was present during the most active phase of AR 11302 with 4 M-flares within less than 3.5 hr. The disappearance of UF 1 was gradual as can be seen in Fig.~\ref{temp_filam}, and therefore probably not directly correlated with the single M-flare occurring right at that time.

\section{Chromosphere above umbral filaments}
   \begin{figure} 
   \centering
    \includegraphics[width=0.5\textwidth]{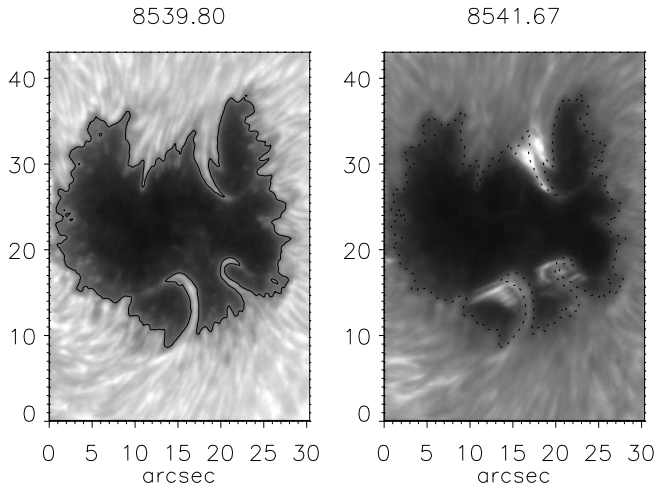}
   \caption{Left: Intensity image of the far blue wing in the 8542 \AA\ line. Right: Intensity image at 8541.7 \AA, showing emission above the UFs, which is spatially shifted towards the direction of the coronal loops. The contours are the same in both images.
                    }
         \label{chromfilam}
   \end{figure}
\begin{figure*} 
   \centering
    \includegraphics[width=\textwidth]{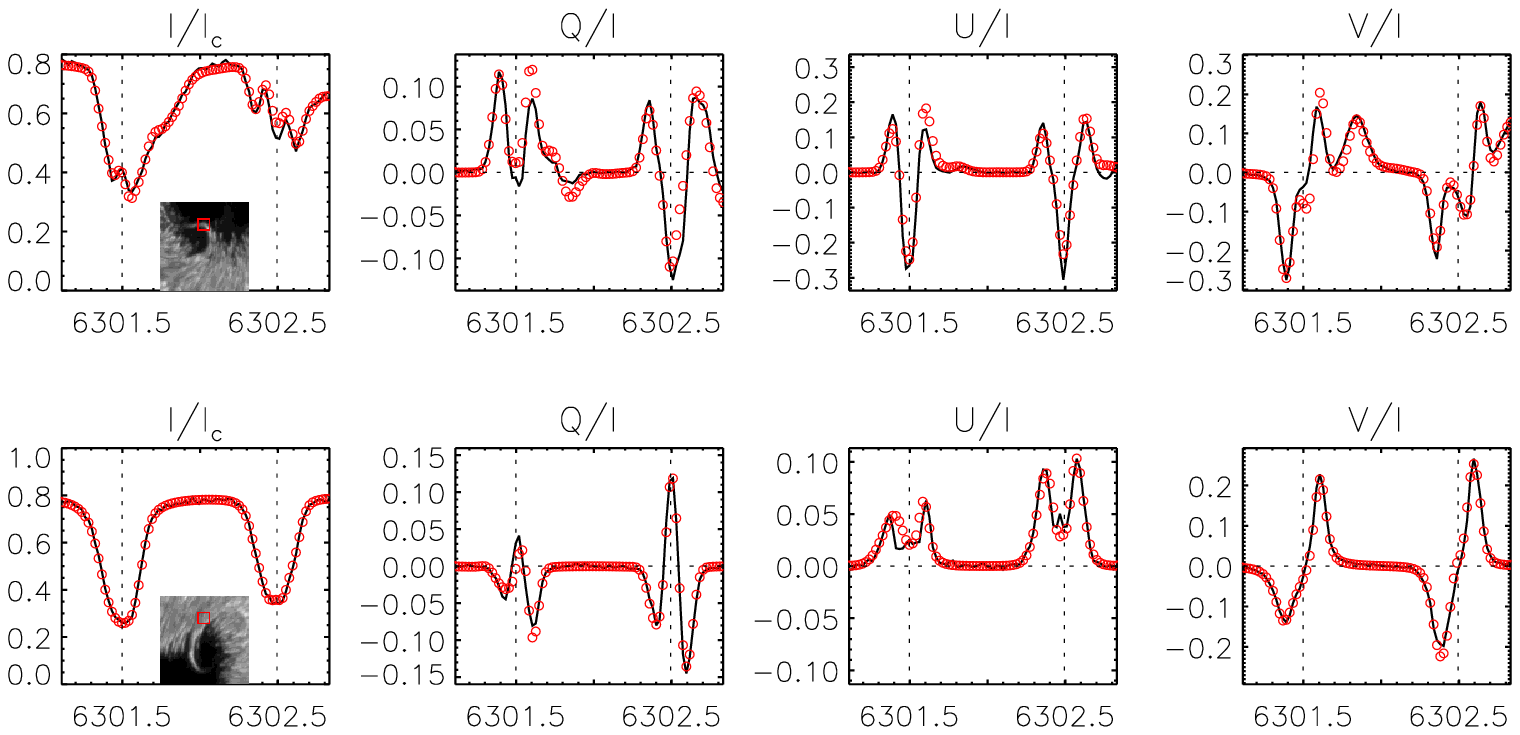}
   \caption{Examples of fits derived from the inversions (red circles) to the Hinode observations (black solid lines). The top profiles show a pixel in the south-eastern UF~1, marked by an enlarged red box in the image. An extra lobe is visible in $V/I$ and the linear polarization states are asymmetric. The bottom profiles show a pixel in the penumbral extension of UF~3 with asymmetries in all Stokes parameters.
                    }
         \label{sirexample}
   \end{figure*}

While the photosphere shows the three filaments as separate features with different life times, at least two of them (UF~1 and 3) seem to be connected in the chromosphere. This connection could be because of a current sheet or low lying loops. Figure~\ref{ibis8542} shows the chromospheric Stokes $I$ observed with IBIS at four different time steps (left to right, UT time given above images) in the 8542 \AA\ line core -- 130 m\AA. An M2.8 flare started at 17.59 UT, reaching a maximum at 18.05 UT, and one of its footpoints is visible in the third and fourth image. The UFs in the chromosphere were bright already before the flare started. The brightness of the northern filament (UF~3) increased significantly after the flare. From SDO/AIA images (Fig.~\ref{aia}) it is evident that the UFs were already bright as soon as they started to be visible in the umbra, indicating that their dissipation of energy might be related to currents because of shearing and not primarily flaring. Flares increase the energy and temperature along these filaments when particles travel along coronal loops to the chromosphere, probably leading to the additional brightening seen in the fourth image. The figure also shows that at least the northern filament extends far outside of the penumbra (arrows in third image) and seems to either lie above the usual penumbral structure or have a higher density. The structure of the connection of the filaments by a bright ribbon crossing the umbra seems to remain constant during at least 25 minutes, with only the intensity changing because of the flare.

 Images in the blue wing of the 8542 \AA\ line also show emission above the UFs, although it can be seen in Fig.~\ref{chromfilam} that the location of the emission is spatially shifted. The two filaments on the southern side of the spot show the emission shifted towards the east (left in image), while the northern filament has emission toward its north-east. By comparing with Fig.~\ref{aia}, it seems that the emission is in the direction of the coronal loops and we can conclude from the right panel of Fig.~\ref{chromfilam} that the energy, which leads to emission, is not dissipated in the lower photosphere, otherwise there would not be an offset between enhanced emission and filament location. Instead, the dissipation seems to happen in the upper photospheric or lower chromospheric level.

\section{Inversions of the Stokes profiles}\label{inversions}
\subsection{Photospheric Inversions}
\subsubsection{Strategy}

By perturbing a model atmosphere iteratively and calculating sets of Stokes profiles, one can find the best fit to the measured Stokes profiles. This fitting process is called inversion and can derive atmospheric parameters for each pixel of the observations. Milne--Eddington (ME) inversions \citep{skumanichlites1987, borreroetal2011} assume no variation with height of atmospheric parameters, making the Stokes profiles look perfectly symmetric. Our UFs however, often show complex Stokes $V$ profiles with three lobes or asymmetries in all Stokes parameters and therefore the ``standard'' ME inversion of Hinode data (MERLIN code) cannot be used for them. Instead, we use the Stokes Inversion based on Response functions (SIR) \citep{ruizcobodeltoro1992}, which allows to specify height-dependent parameters and even two magnetic atmospheric components.

It is known that the penumbral atmosphere
can be explained in terms of two components \citep{Sol93, Mar00, Bor02, Schl02, Bel04}. One component is more vertical
and has a stronger magnetic field and it is associated with the background magnetic field in the sunspot. 
The more horizontal and weaker magnetic field component is related to the flux tubes, i.e.
the filamentary penumbra. These two components are an idealized
representation of the sunspot atmosphere that help us to get a better
understanding of the results of the inversion. Although they have been referenced with different terminologies 
\citep{Lit93, Sol93, Tit93}, the picture is nearly equal the one described above. As explained further below, we will use two atmospheric components for the inversions for the same reason.

To determine the best strategy for the inversions, we selected several pixels 
along the filaments for testing. They show asymmetric Stokes profiles, not 
only in Stokes $V$, but also in $Q$ and $U$, several lobes in the same wing of 
the Stokes $V$ profile or asymmetric Stokes $I$ profiles.
Figure~\ref{sirexample} shows two sets of these profiles (solid black) with 
their location near the filament indicated by the red box in the small image. 
The inversions are overplotted with red circles.
In the first row, Stokes V shows three lobes. A simple way to explain these
kind of profiles is by considering the presence of two components (or atmospheres) 
in the same pixel with opposite polarities and different velocities.
In the second row, all Stokes profiles are asymmetric. 
This tells us that these profiles are associated with an atmosphere with 
gradients, possibly in $v_{\rm LOS}$ and $B$.
\begin{figure*}
   \centering
   \includegraphics[width=\textwidth]{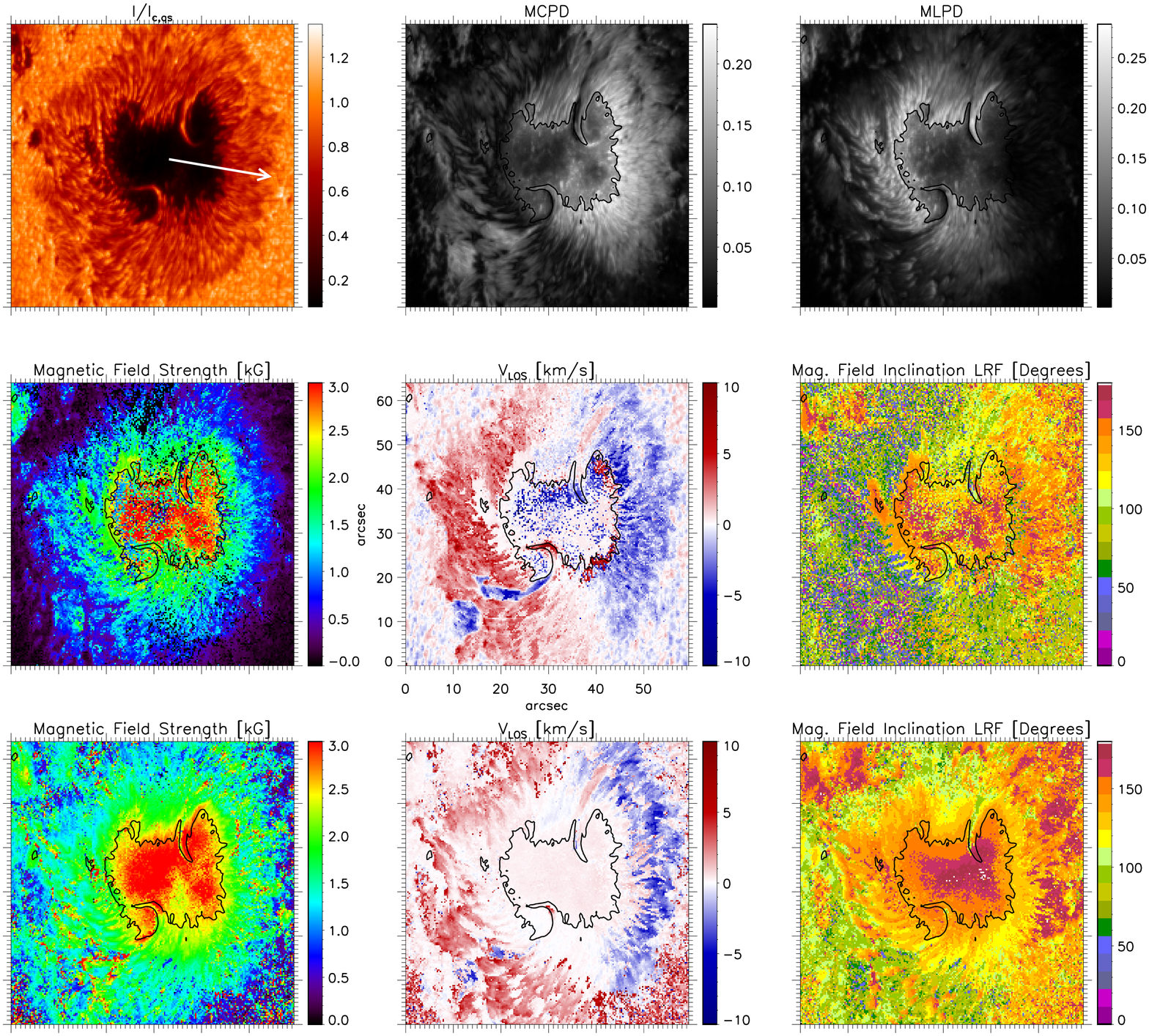} 
   \caption{Top row: intensity, circular and linear polarization degree of the Hinode map taken on Sep 24, 2011 with timestamp 13:00:05. The arrow points towards disk center. Middle and bottom row: SIR inversion results for model 1 and 2, respectively. The magnetic field, the line-of-sight velocity and the inclination in the local frame are shown. Note that there is a flow opposite to the regular Evershed flow for both filaments.}
   \label{invoverview}
\end{figure*}

The mean circular polarization degree (MCPD) and mean linear polarization
degree (MLPD) also give us valuable information about the atmosphere where
the Stokes profiles were generated.
The MCPD is calculated as 
\begin{equation}
{\rm MCPD} =
(\int_{\lambda_{0}}^{\lambda_{1}}{\frac{|V(\lambda)|d\lambda}{I(\lambda)}})
        /(\lambda_{1} - \lambda_{0}).
        \end{equation}
The MLPD is calculated as 
\begin{equation}
{\rm MLPD} = (\int_{\lambda_{0}}^{\lambda_{1}}{\frac{\sqrt{Q^2(\lambda) + 
        U^2(\lambda)}}{I(\lambda)}d\lambda})/(\lambda_{1} - \lambda_{0}).
        \end{equation}
        
The integrals are evaluated from  $\lambda_{0}=6301.29$~\AA\ to
$\lambda_{1}=6301.71$~\AA\ in the case of
 \ion{Fe}{1} 6301 \AA, and from $\lambda_{0} = 6302.27$~\AA\
 to $\lambda_{1}=6302.70$~\AA\ for \ion{Fe}{1} 6302 \AA.
These wavelength intervals cover all the relevant polarization signals,
avoiding the contribution of the continuum.
The MCPD and MLPD are good proxies of the vertical and horizontal component of the vector magnetic
field, respectively.

Figure \ref{invoverview} shows the parameters corresponding to the data observed
 on Sep 24, 2011 at 13:00. Top row, left to right, are: the slit-reconstructed intensity map (normalized with
respect to the continuum in an average quiet Sun profile), the MCPD and
MLPD maps. In the MCPD and MLPD maps we can distinguish two parts. In the
MCPD map the limb-side (left side on the image) has lower values (darker) than the 
center-side (right). This effect is opposite in the MLPD map. \cite{Schl02} and \cite{Bel04} described this bimodal behavior of the polarization degree maps in the penumbra as
the co-existence of two different magnetic components in the penumbra. We can
also observe this sharp variation of the polarization in both UFs. UF~1 and UF~3 are black at the limb-side
in both polarization degree maps. On the center-side, UF~3 is gray in the MCPD map and white in the MLPD, just outside of the 
contour in both maps and UF~1 looks similar. The interpretation given by \citep{Bel04,
Bor04}, the observed Stokes profiles, and the model of the penumbra discussed above suggest to use a two component model atmosphere for our inversions.

The number of nodes of the atmospheric parameters (degrees of freedom) in the inversions used in this paper 
are shown in Table \ref{table_inversion}. The quadratic behavior of the magnetic field strength and velocity in the line of sight (LOS; nodes = 2) allow a gradient in these parameters. Similarly, the larger 
degrees of freedom in the temperature allow a good fit to the (often asymmetric) Stokes $I$
profiles. In addition, we included a variable 
stray light component, which is introduced as an average intensity profile in a 
magnetically quiet area located in the vicinity of the sunspot. 
Although other combinations of nodes were tested, the selected 
ones offer a good trade-off between reliability of the output atmosphere and the errors
given by the inversion code on one hand and the fits of the Stokes
profiles on the other. The difference between the two atmospheric components in
each pixel of the data was the initial inclination (0$^\circ$ and 45$^\circ$ in the LOS frame) and
the initial magnetic field strength (1kG and 2 kG). These values were deliberately chosen not to match the expected solution, to allow the inversion code to sample a large parameter space.  
\begin{table} 
\caption{Number of nodes used in the inversions.\label{table_inversion}}
\begin{tabular}{c|c|c} 
\hline 
Physical Parameter & Cycle 1 & Cycle 2\\
\hline\hline
Temperature  & 2 & 4 \\
LOS Velocity ($v_{\rm LOS}$)& 1 & 2 \\
Magnetic Field Strength ($B$)  & 1 & 2 \\
Magnetic Field Inclination & 1 & 1 \\
Magnetic Field Azimuth & 1 & 1 \\
\end{tabular}
\end{table}

\subsubsection{Analysis and Results}

Table \ref{tab3} shows the dates, location, and area of the
nine inverted maps. The first time in the table is the timestamp in the Hinode database (start obs). 
The time in parenthesis refers to the time when the spectrograph slit was located approximately in the middle 
of the studied sunspot and this was our reference time for the location and $\mu$. The last column shows
the inverted area for each scan. The nine maps were observed far away from disk center. 
Therefore, their $v_{\rm LOS}$ maps show a clear Evershed flow pattern. This pattern can be clearly seen in the two atmospheric components (\textit{model 1} and \textit{2}) in the middle and bottom rows of Fig.~\ref{invoverview}. These rows show from left to right: $B$, $v_{\rm LOS}$\footnote{We used the averaged value of the $v_{LOS}$ 
in a dark area of the umbra (9\arcsec $\times$ 9\arcsec) to calibrate the $v_{LOS}$ zero value \citep{beckers1977,rimmele1994}.} and the magnetic field inclination in the local reference frame. The intensity contour of the umbra-penumbra edge was overplotted in all the parameters, except for the intensity map. Two UFs are visible in the lower and the upper part of the umbra. Keeping the numbering from the previous sections, the southern (lower one) is called UF~1 and the northern UF~3.  

The magnetic field strength for model~1 is lower than for model~2. The magnetic field of the UFs is closer to umbral than to penumbral values. The inclination of model 1 is more horizontal (closer to the solar surface). Model 2 is associated with a wide range of inclination values, from nearly vertical (with respect to the vertical in the local reference frame) in the umbra to nearly horizontal in the outer part of the penumbra. 
  \begin{figure*}
   \centering
    \includegraphics[width=\textwidth]{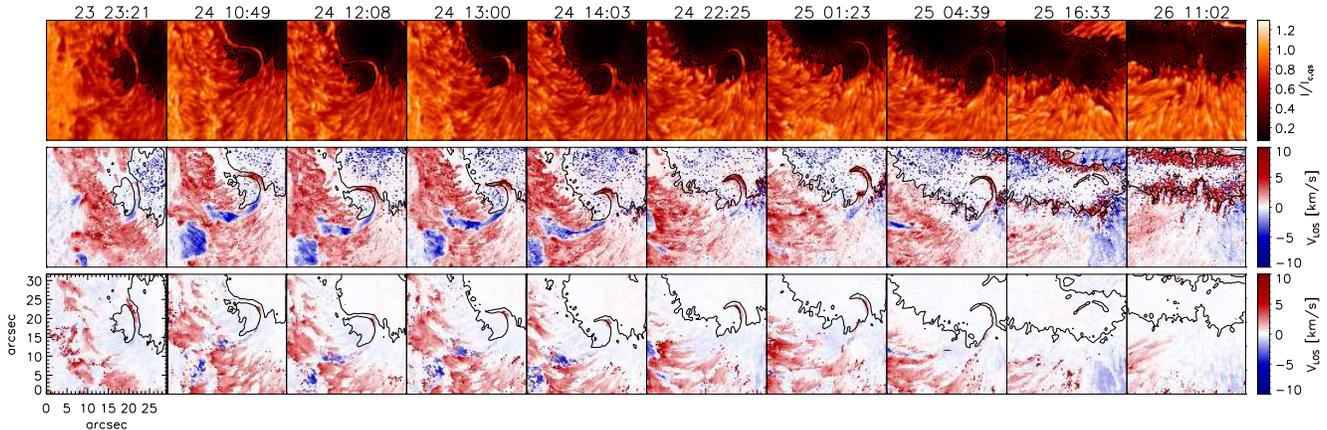} 
   \caption{Temporal evolution of the intensity (top row) and $v_{\rm LOS}$ for models 1 and 2 (middle and bottom row, respectively). A reversed Evershed flow is clearly visible, mostly in the more horizontal model 1, but also at the footpoints of UF~1 in model 2.
                    }
         \label{doppevol}
   \end{figure*}

The average filling factor of model~1 is $30\%\pm5\%$ in the inner penumbra
and $60\%\pm5\%$ in the outer penumbra. Obviously, model~2 complements these values to 100\%. The average atmospheric filling factor of UF~3 is (30\%,70\%)$\pm$5\% and for UF~1 (50\%,50\%)$\pm$5\% for (model 1, model 2). The stray light contribution is less than $5.0\%\pm0.5\%$ for the penumbra. 

\begin{table}
\caption{List of maps observed by Hinode-SOT/SP used for the study of the
temporal evolution of the UFs. \label{tab3}}
\begin{tabular}{lccc}
\hline
Date & Location&{$\mu=\cos\theta$}  & Inv.~Area ($\arcsec\times\arcsec$) \\
\hline\hline
2011.09.23 23:21 (23:56) &  N12E56 & 0.56 & $59\times64$ \\
2011.09.24 10:49 (11:19) &  N12E51 & 0.64 & $59\times64$ \\
2011.09.24 12:08 (12:30) &  N12E50 & 0.65 & $59\times64$ \\
2011.09.24 13:00 (13:30) &  N12E50 & 0.66 & $59\times64$ \\
2011.09.24 14:03 (14:30) &  N12E49 & 0.66 & $42\times64$ \\
2011.09.24 22:25 (22:56) &  N12E44 & 0.72 & $59\times64$ \\
2011.09.25 01:23 (01:54) &  N12E43 & 0.74 & $59\times64$ \\
2011.09.25 04:39 (05:10) &  N12E41 & 0.76 & $59\times64$ \\
2011.09.25 16:33 (17:06) &  N12E35 & 0.82 & $59\times64$ \\
2011.09.26 11:02 (11:36) &  N12E25 & 0.91 & $71\times77$ \\
\end{tabular}
\end{table}

The Evershed flow is moving outwards from the umbra to the outer part of the penumbra.  If the sunspot is located 
off disk center, the component of the velocity vector in the LOS will be redshifted on the limb-side of the sunspot and blueshifted on the center side. 
However, the UFs show opposite colors. In model 1, UF~1 is blueshifted 
with respect to the LOS from the mid penumbra and redshifted when it reaches 
the umbra. Its surroundings are clearly redshifted, as it is on the limb-side 
of the penumbra. UF~3 is showing the same behavior on the center-side with 
opposite values: it is redshifted in the mid penumbra and blueshifted inside 
the umbra, while the penumbra is blueshifted. This means that the flow 
associated with the UFs is directed from the middle part of the penumbra to the 
mid umbra, opposite to the Evershed flow.  If we assume that the magnetic 
field is driven by the kinetic flow, which is usually true in the photosphere, 
the observed UFs move magnetic flux into the umbra.  Model 2 only 
shows a contribution in the footpoint of UF~1, with a blueshifted patch 
in the mid penumbra and a redshifted patch in the mid umbra. UF~3 in 
model 2 is hardly showing a blueshifted footpoint in the umbra at the 
coordinates [36,39], but almost the whole part of UF~3 is redshifted.  

In summary, we can distinguish a mostly horizontal, weaker magnetic component (model 1) that is 
mainly driving the Evershed flow, but the UFs are clearly showing a flow 
opposite to the Evershed flow. A stronger background component (model 2) shows a large variation of its inclination, 
going from almost vertical in the inner penumbra to almost horizontal in the
outer penumbra. The UFs are clearly associated with the horizontal component, but their footpoints are  
visible in the background component. We will discuss the topology of the UFs in Section \ref{sec_dis}.

Figure~\ref{doppevol} shows the temporal evolution of $v_{\rm LOS}$ of UF~1. 
The day (all images are from Sep 2011) and the hour of the inverted data are displayed above the intensity images, in the first row of the figure. The second and third rows show $v_{\rm LOS}$ for model 1 and 2, respectively. The umbra-penumbra intensity contour was overplotted for both models. UF~1 splits the umbra into two parts between the first snapshot (23 23:21) and the third one (24 12:08), but it does not resemble typical LBs: Its southern end is very thin and does not have a granular structure, while the northern part of the structure does show a granular pattern. UF~1 seems to form from only the filamentary part. The granular part seems to merge with the penumbra.
At this time, the sunspot is close to the limb ($\mu = 0.56-0.65$), and therefore we cannot discard some projection effects. 
We cannot conclude whether UF~1 is the final stage of a previous LB, a failed LB formation, or a different phenomenon. 
In these three first snapshots, the UF starts to develop a prominent blueshifted 
footpoint. The velocity in the central part of the footpoint is as large as  $-9$ km/s. This velocity goes toward zero closer to the umbra, where the other footpoint shows redshifted velocities.  In these early stages, $v_{\rm LOS}$ reaches the
largest blueshifted values in model 1. In the fourth and fifth snapshots,
the UF becomes thinner, but it is always showing a blueshifted footpoint in the penumbra and a redshifted footpoint in the umbra, while the Evershed flow is clearly redshifted in the limb-side of the penumbra. Its maximum 
length is $\sim$17\arcsec\ in the blueshifted part, and $10\arcsec$ taking 
into account the non-shifted (displayed in white color) and redshifted part.
UF~3 shows a similar behavior with a smaller values of $v_{\rm LOS}$. It may be because of a real smaller velocity or because of a larger uncertainty to determine the values of the parameters in the center-side of the penumbra \citep{Bor04}.

  \begin{figure*}
   \centering
    \includegraphics[width=.47\textwidth]{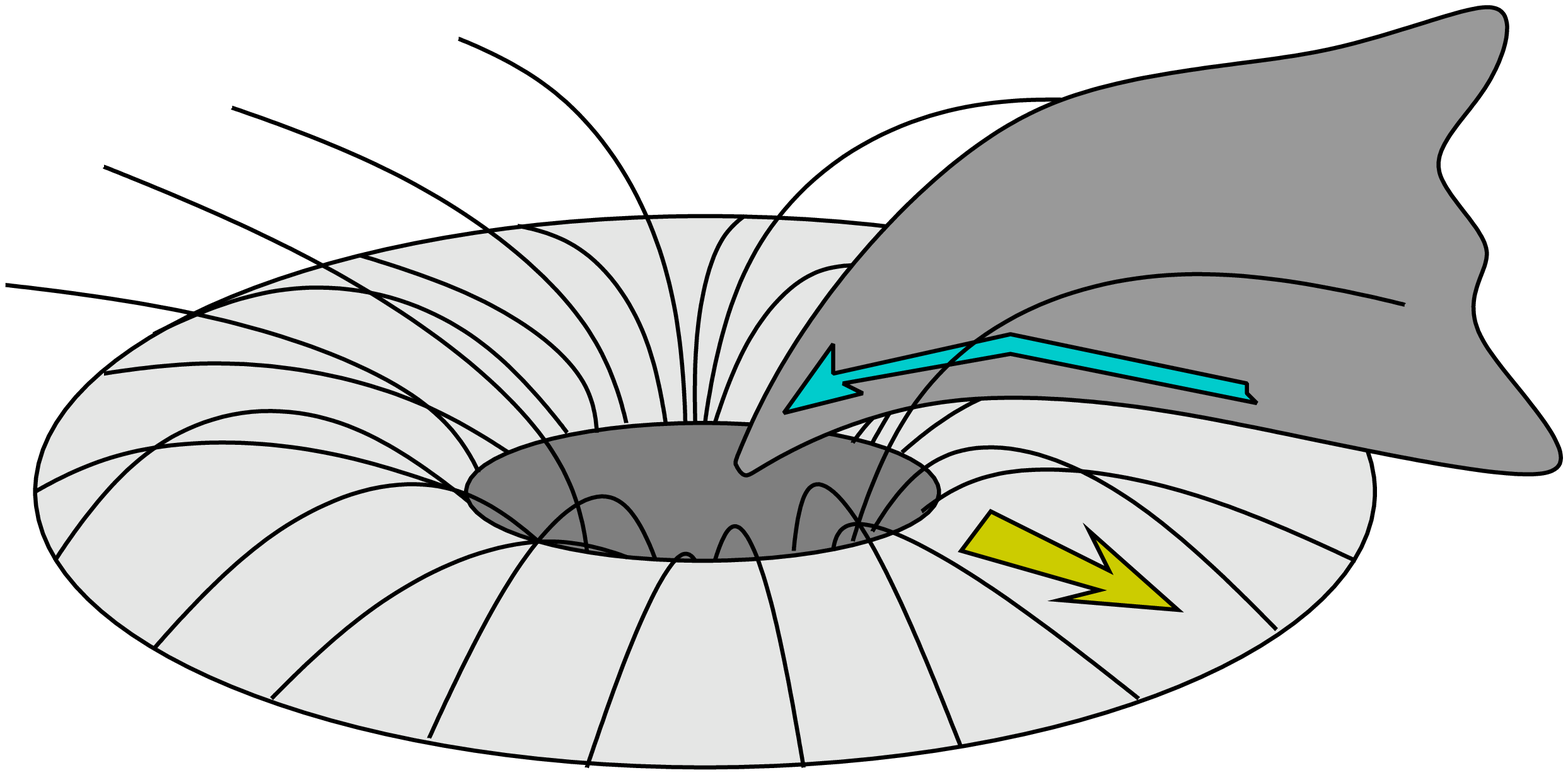}
      \includegraphics[width=.45\textwidth]{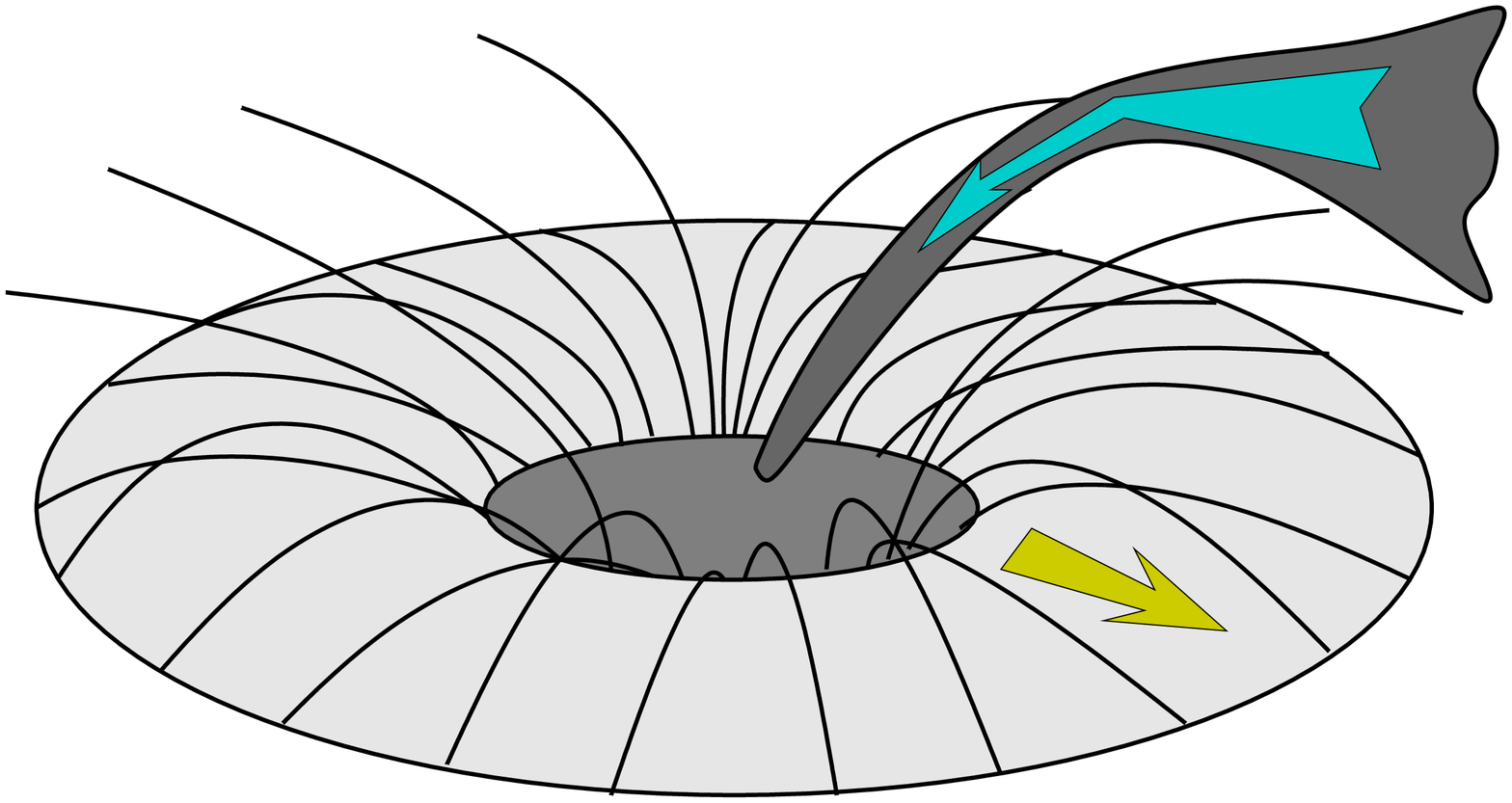}
     \caption{Two simple schematics depicting a sunspot, which might explain the observed umbral filaments. The model on the left shows the filament (gray) as a sheet, spanning many atmospheric heights. On the right, the UF is shown as dense structure, which might block the observer from seeing the material below it.
                    }
         \label{xfig}
   \end{figure*}







\section{Discussion}\label{sec_dis}
To understand what UFs are, we need to address several questions about their inverse Evershed flow and their morphology and evolution. In the following, we will present two schematic scenarios, shown in Fig.~\ref{xfig}, which we will then use to explain some of the observed phenomena.

Each part of Fig.~\ref{xfig} shows a simple sunspot model, where the more horizontal 
 component of the penumbral magnetic field is represented by lines connecting 
 the outer part of the umbra with the outer part of the penumbra, while the more vertical or background component is displayed as unconnected lines
 going up from the outer part of the umbra. The model on the left will be called the {\it umbral filament sheet model} and on the right, we show the {\it massive umbral filament model}. 

In the sheet model, the UF is formed by a sheet, which cuts the solar atmosphere from the photosphere to the 
 corona. This model can be explained as a 2D extension of a bifurcation of field lines through the solar 
atmosphere  \citep[see Figure~7.22 in][]{parker1994}. Thus, our sheet might be a topological feature of
separating flux bundles that are adjacent to each other in and above the umbra and connecting to spatially separated regions,
for instance in the network. The base of the sheet is as curved as the horizontal component of
 the penumbra. Mass is flowing {\it inside} the sheet, from outside of the sunspot to
 the umbra. This inflow would be a {\it siphon flow} produced by the pressure difference between the umbra base and the 
magnetically connected far regions in the network. There could be shearing at the boundary layers, leading to energy dissipation, which might explain the emission at chromospheric and upper photospheric heights. The sheet would need to be dense enough to keep its structure for a few days, but its mass cannot be too large to split the sunspot (in case of UF 1 and UF 2) or just enough to do so (case of UF~3). The regular Evershed flow may just surround the sheet, but it is unclear what would happen at the boundary. 
   
Another possibility is that we are observing a thick, magnetic flux tube, with 
 a high enough density that the observed region (i.e., $\tau=1$) is formed higher for
 the tube than for the penumbra, as depicted in the massive umbral filament model. In this case, the opposite flow may be the usual reversed Evershed flow, which is commonly observed higher in the solar atmosphere \citep{stjohn1913, beckers1962,maltby1975,solanki2003}. The regular Evershed flow may be present below our flux tube.   
 However, it might be difficult to explain how this elevated, massive flux tube evolves into a LB or a structure similar to a LB, as it seems to happen
 with UF~3, which separates the umbra into two parts at some point during its lifetime. One possibility is that this flux tube becomes heavier and heavier, and falls down to the photosphere, with enough mass and energy to split the sunspot, which however is the opposite picture to regular convective motions that form LBs.

Which model can now explain the observations better? The reversed Evershed flow can be explained by both, either as a pressure-driven flow inside the sheet or as a regular inverse flow that is commonly observed in the chromosphere. The boundary between regular and opposite Evershed flow is harder to explain in the sheet model, because one would expect turbulent motions, which are not observed (with our spatial resolution). Both models could explain the visibility of the UFs in different atmospheric layers, but the data do not allow to determine if the inflow is present at all heights. Influences on coronal loops through photospheric motions may also be explained by both models, thus no model excludes or directly supports a relation to flares. The motions, especially the curling and unwinding might be explained by the apparent rotation of the sunspot, but calculations would be needed to determine if the massive flux tube, being located higher in the atmosphere, would easily follow these photospheric motions. Both models could support the apparent connection of two UFs in the chromosphere, either as connected sheets or as density variations within the massive flux tube. In summary, both models may be used to explain the observed features and without additional observational data we cannot exclude either of them. High-sensitivity chromospheric spectropolarimetric data would be needed to for example determine the inclination of the magnetic field. It is unclear how the inclination varies from close to horizontal in the photosphere to probably vertical in the lower corona, assuming that the field is oriented along the observed coronal loops. These simple models also cannot explain under which solar conditions UF may form.

For the global picture, we can assume that there is significant shearing through the observed photospheric motions. This will probably lead to ${\rm rot}(B) \ne 0$ in the higher atmospheric layers. Currents are thus induced, and energy is dissipated, which is probably related to the observed emission near the UFs at chromospheric heights. The emission was already present before the M-flare started and increased right after the flare. With a radiative cooling time of about 90 s at this height, energy needs to be supplied continuously to keep the ribbon-like structure bright for a longer period of time.





\section{Conclusions}
The different instruments used in this study of AR 11302 allow us to determine a more complete picture through different atmospheric layers. The UFs are observed in the photosphere as thin, curled filaments, which reach from the penumbra well into the umbra. UF~1 seems to evolve from an atypical LB and rotates around the sunspot's umbra; UF~2 is launched very fast into the umbra, has the shape of a hook and does not split the umbra and it disappears within 8.5 hr; UF~3 seems to evolve from a looped penumbral filament, whose footpoints first stay in the penumbra, but then one of them seems to move, creating the actual UF. UF~3 seems to evolve into a LB, which only separates a very small part of the umbra, but this part disappears after a few days. 

 The lifetime of the UFs varies from several hours to a few days: UF~1 approximately lived for 3 days, UF~2 for 0.4 days and UF~3 lived as thin, curled structure for 1 day, then it existed for two more days as a LB. Their presence correlates with the most flare-productive phase of AR 11302.
 
These UFs are not a common occurrence in sunspots and to our knowledge, have not been reported before. Their most striking property is a flow opposite to the Evershed flow, which goes from the outer penumbra into the umbra and is well visible in the inverted photospheric data. Plasma is therefore probably transported from outside of the sunspot into the umbra, assuming that the flow follows the inclined field lines.

The UFs are also visible in chromosphere and in the corona as extended structures going from the
 umbra to the network, i.e., further outside of the penumbra. The chromospheric images show emission at the location of these filaments and coronal images allows us to identify them as footpoints of bright coronal loops. This fact, combined with their dynamic evolution can lead to the speculation that these photospheric flows influence the structure of the overlying coronal loops, leading to flares as the loops rearrange to a more potential state.

\acknowledgments
We are very grateful to BC Low for his valuable comments. We also thank R. Schlichenmaier, P. Judge, and B. Ruiz Cobo for helpful suggestions. IBIS is a project of INAF/OAA with additional contributions from Univ.~of Florence and Rome and NSO. \textit{Hinode} is a Japanese mission developed and launched by ISAS/JAXA, with NAOJ as domestic partner and NASA and STFC (UK) as international partners. It is operated by these agencies in co-operation with ESA and NSC (Norway).

\bibliographystyle{apj}
\bibliography{journals,ibisflare,flare_asd}

\end{document}